# Superconductivity at 2.2K in $La_3Ni_4P_4O_2$

T. Klimczuk[1,2],* T.M. McQueen[3],* A.J. Williams[3], Q. Huang[4], F. Ronning[1], E.D. Bauer[1], J.D. Thompson[1], M.A. Green[4,5], and R.J. Cava[3]

[1] Los Alamos National Laboratory, Los Alamos, NM 87545, USA

[2] Faculty of Applied Physics and Mathematics, Gdansk University of Technology, Narutowicza 11/12, 80-952 Gdansk, Poland

[3] Department of Chemistry, Princeton University, Princeton NJ 08544

[4] NIST Center for Neutron Research, National Institute of Standards and Technology, Gaithersburg MD 20899

[5] Department of Materials Science and Engineering, University of Maryland, College Park, Maryland 20742-2115, USA

* These authors contributed equally to this work.

**Abstract**

We report the observation of superconductivity in $La_3Ni_4P_4O_2$ at 2.2 K. The layer stacking in this compound results in an asymmetric distribution of charge reservoir layers around the $Ni_2P_2$ planes. The estimated Wilson ratio, $R_w \approx 5$, indicates the presence of strongly enhanced normal state susceptibility, but many of the basic superconducting characteristics are conventional. The estimated electronic contribution to the specific heat, $\gamma \approx 6.2$ mJ mol-$Ni^{-1}$ $K^{-2}$, is about 2/3 of that found in layered nickel borocarbide superconductors.

Like the high $T_c$ copper oxides[1], the oxypnictide superconductors[2,3,4] are based on square metal layers. Several structural classes of these superconductors (with X-$M_2$-X layers, where M = Fe or Ni, and X = P, As, or Se), are currently known. The simplest is based on β-FeSe[5], which consists of Se-$Fe_2$-Se layers only, and can be considered as the parent compound of the family. A related compound, $Li_xFeAs$, also superconducts[6,7]. Others, typified by $BaNi_2P_2$[8] and K-doped $BaFe_2As_2$[9], are derived from the β-FeSe-like parent by inserting layers of alkali earth ions between X-$M_2$-X layers. Superconducting LaFeAsO and LaNiPO[2,3] are derived instead by inserting whole La-$O_2$-La layers between the X-$M_2$-X layers. Here we show that $La_3Ni_4P_4O_2$, a compound that has a mixed-layer structure between $LaNi_2P_2$ and LaNiPO, is superconducting with a $T_c$ of 2.2 K. The stacking of inequivalent charge reservoir layers in $La_3Ni_4P_4O_2$ leads to an asymmetric distribution of ionic layers around the $Ni_2P_2$ planes, a structure-derived electronic feature that is not observed in any other known member of the superconducting oxypnictides.

Polycrystalline $La_3Ni_4P_4O_2$ samples were prepared from $Ni_5P_4$, elemental P, and dry $La_2O_3$[10]. Physical property characterization was performed on a Quantum Design PPMS. Neutron diffraction data were collected at the NIST Center for Neutron Research on the BT-1 powder neutron diffractometer[11]. The crystal structure was determined by a Rietveld fit[12,13] to powder neutron diffraction data. The diffracted peak positions and intensities were well fit by a tetragonal $Pr_3Cu_4P_4O_2$-type crystal structure[14]. The fractional occupancies of each crystallographic site were independently refined and, within two standard deviations, were all found to be equal to one. Therefore the structure can be considered stoichiometric, and for the final refinement, all site occupancies were fixed at unity. The refined structural parameters are summarized in Table I. Small quantities of two impurity phases are present in the best preparations; 10.0(2) % of LaNiPO and 6.1(2) % $LaNi_5P_3$ were included in the final fit for the sample characterized in this study.

The crystal structure of $La_3Ni_4P_4O_2$ is made from covalently bonded P-$Ni_2$-P layers, the common structural motif seen throughout the superconducting oxypnictide family. The M-X bond lengths and M-X-M bond angles within these covalent layers are under discussion as important chemical influences on the electronic properties in this family of compounds (see e.g. ref. 15), as is X-X bonding[16]. These layers are separated in the new compound by an alternating pattern of layers of La, and La-$O_2$-La. Figure 1 shows the crystal structure, which is an ordered structural mixture of LaNiPO[3] and $LaNi_2P_2$[17], and its relationship to those phases and the parent β-FeSe structure type. These compounds form a homologous series with the formula $(La_2O_2)_m[La/Ba]_n(Ni_2P_2)_{n+m}$, and, including $La_3Ni_4P_4O_2$, the m,n pairs (1,0), (0,1) and (1,1) are known to support superconductivity. In Figure 1, the $La^{3+}$ and $O^{-2}$ ions are drawn using ionic radii and the M and X atoms are drawn using covalent radii. This rendering reflects a chemically interesting aspect of the oxypnictide superconductors – the intimate intergrowth of highly ionic layers such as $La^{3+}$ and $(La_2O_2)^{2+}$ with highly covalent $X-M_2-X$ layers such as $Ni_2P_2$ or $Fe_2As_2$. These ionic layers are the "charge reservoirs"; completely analogous to those in the layered high Tc copper oxides. What is unique about $La_3Ni_4P_4O_2$ among the oxypnictide superconductors is that the distribution of charge reservoir layers around the $Ni_2P_2$ plane is highly asymmetric; i.e. the sequence of layers is $[La^{3+}]-[P-Ni_2-P]-[La-O_2-La]^{2+}$. Thus, the $Ni_2P_2$ layer is sandwiched between two different ionic layers – an arrangement that should lead to the presence of a local electric dipole field across the superconducting plane.

Figure 2 shows the superconducting transition for $La_3Ni_4P_4O_2$ characterized by magnetic susceptibility ($H_{DC}$ = 10 Oe). The critical temperature ($T_c$) is about 2.2 K. The minor diamagnetism visible in the temperature range 2.5 K to 4 K is associated with the presence of the 10 % LaNiPO impurity phase. As is the usual case, pinning of vortices produces a smaller field-cooled (FC) signal compared to zero-field-cooling (ZFC). The diamagnetic response in ZFC measurements is substantially larger than $1/4\pi$, which we attribute to demagnetization effects of the rectangularly shaped sample. To estimate the demagnetization factor (d), measurements of the low field magnetization were made at 1.9 K, shown in Figure 3. Assuming that the initial linear response to fields less than 50 Oe is perfectly diamagnetic, that is, that the slope $dM/dH$ is $-1/4\pi$, we obtain a demagnetization factor that is consistent with the sample's shape. Applying this demagnetization factor gives the maximum ZFC susceptibility in Figure 2 of approximately $-1.1(1/4\pi)$. As shown in the inset of Figure 3, M(1.9 K) begins to deviate from a fitted linear dependence on H at a field $H^* \approx 60$ Oe, giving a lower critical field[18], taking into account the demagnetization factor, of $H_{C1}(1.9K) = H^*/(1-d) \approx 90\pm10$ Oe. The estimated uncertainty reflects criteria that H* is either the field above which M first deviates from the fitted line or the field where M deviates by 2.5 % above the fitted curve. On the basis of this $H_{C1}$ at 1.9 K and $T_c$ in zero field, a crude estimate of the zero-temperature lower critical field is about 520 Oe, which would imply a Ginzburg-Landau superconducting penetration depth of approximately $\lambda_{GL}$ of 550 $\pm$ 50 Å[18].

The inset of Figure 2 shows the temperature dependence of the estimated normal-state magnetic susceptibility. Low field (H = 1 kOe) measurements indicated the presence of small amounts of ferromagnetic impurity with a low coercive field, likely elemental nickel, in the best samples. This makes the usual low-field measurement of susceptibility invalid due to the relatively high contribution of the ferromagnetic impurity to the measured magnetization at low fields. Therefore, to approximate the intrinsic susceptibility of $La_3Ni_4P_4O_2$, the difference in magnetization ΔM between applied fields of 50 kOe and 40 kOe, well above the coercive field of the ferromagnetic impurity, was employed to estimate the susceptibility χ=ΔM/ΔH. Above 250 K $La_3Ni_4P_4O_2$ exhibits temperature independent Pauli paramagnetism with $χ_0 ≈ 1.8·10^{-4}$ emu mol$^{-1}$. Below 250 K, χ(T) increases with decreasing temperature and reaches maximum at about 14 K. This may reflect the existence of a magnetic transition in this phase, but the presence of LaNiPO and $LaNi_5P_3$ impurities precludes more detailed analysis at this time.

Specific heat ($C_p$) measurements performed at 0 and 90 kOe magnetic fields are shown in Figure 4. The data in zero field (closed circles) show an anomaly at $T_c$ = 2.3 K, close to the $T_c$ determined by susceptibility measurement. (There is also a small anomaly near 4 K due to the LaNiPO impurity.) The normal-state specific heat measured at $μ_0H$ = 9 T, shown by open circles, was fitted to $C_p = γT+βT^3+ A/T^2$, where the first, second and third terms represent electronic, lattice, and nuclear Schottky contributions, respectively. The fit to these data allows us to estimate γ = 6.2 mJ mol-Ni$^{-1}$ K$^{-2}$, β = 0.14 mJ mol-Ni$^{-1}$ K$^{-4}$, and A = 0.49 mJ mol-Ni$^{-1}$ K. The value of γ is smaller than those found in the layered nickel borocarbide superconductors, e.g. $LuNi_2B_2C$ (8.9 mJ mol-Ni$^{-1}$ K$^{-2}$)[19], which also have significantly higher $T_c$s (16.5 K for $LuNi_2B_2C$). Given both γ and $χ_0$, the Wilson ratio can be estimated for $La_3Ni_4P_4O_2$, yielding $R_W = \frac{(\pi k_B)^2 χ_0}{3μ_{eff}^2 γ} ≈ 5$ ($μ_{eff}$=2.76*10$^{-21}$ erg/Oe was calculated from the χ(T) fit of 50-350K to Curie-Weiss law)  This is significantly higher than the free electron value ($R_w$=1) and indicates a spin enhanced susceptibility. The normalized specific heat jump at $T_c$ can also be calculated from γ. The experimental value for $La_3Ni_4P_4O_2$, $ΔC_{el}/γT_c$ = 1.25, is lower than the BCS prediction[20] of 1.49, likely due to the influence of the impurity phases on the measured γ. The simple Debye model connects the β coefficient and Debye temperature ($Θ_D$) through $Θ_D = \left(\frac{12π^4}{5β_m}nR\right)^{1/3} ≈$ 360K, where R = 8.314 J mol$^{-1}$ K$^{-1}$, $\beta_m = 4\beta$ = 0.56 mJ mol K$^{-4}$ and n = 13 for La$_3$Ni$_4$P$_4$O$_2$. With this value for $\Theta_D$, we can estimate the electron–phonon coupling constant ($\lambda_{ep}$) from McMillian's relation[21]: $\lambda_{ep} = \dfrac{1.04 + \mu^* \ln(\Theta_D/1.45T_c)}{(1 - 0.62\mu^*)\ln(\Theta_D/1.45T_c) - 1.04}$, where $\mu^*$ is a Coulomb repulsion constant. Taking a typical value for $\mu^*$ = 0.10, we estimate $\lambda_{ep} \approx 0.5$, which implies that La$_3$Ni$_4$P$_4$O$_2$ is a weakly to moderately coupled superconductor, consistent with the specific heat jump at T$_c$.

Electrical resistivity measurements are shown in the main panel of Figure 5. Between room temperature and 5 K, the resistivity decreases by nearly a factor of 10, which is reasonable for a polycrystalline sample. The lower inset shows the resistivity near $T_c$ for representative applied fields. The higher transition is associated with the LaNiPO impurity, which superconducts near 4 K. Due to the presence of this superconducting impurity phase, we use the temperature at which the resistivity reaches half its normal state value, $\rho(T) = 1/2\ \rho_{res}$, to define the mid-point $T_c$ of La$_3$Ni$_4$P$_4$O$_2$. With this definition of $T_c$, the upper inset of Figure 5 presents the temperature dependence of the estimated upper critical field ($H_{C2}$). The initial slope – $dH_{C2}/dT_c$ = 0.31(1) T/K permits an estimate the zero-temperature upper critical field $H_{C2}(0)$ from $0.7T_c\ dH_{C2}/dT_c \approx$ 6 kOe[22]. With this estimate for $H_{C2}(0)$, the zero-temperature superconducting coherence length can be estimated[18] as $\xi_0 = (\phi_o/2\pi H_{C2}(0))^{1/2} \approx 240\pm20$ Å, where $\phi_0$ is the flux quantum. This is a reasonable value, but is significantly larger than that found in the related Ni-based superconductors YNi$_2$B$_2$C (65 Å)[23] and MgCNi$_3$ (47 Å)[24]. The present analysis does not reveal any overtly unconventional behavior in La$_3$Ni$_4$P$_4$O$_2$ beyond the presence of a spin-enhanced susceptibility. While there is widespread belief that there is a relationship between superconductivity and magnetism in the Fe-based oxypnictides, early band structure calculations on LaNiPO[25] have been interpreted as indicating that the nickel oxypnictides are conventional superconductors. The issue of conventional vs. unconventional superconductivity in the Ni-based superconductors – the layered borocarbides and MgCNi$_3$ - continues to be an open one, however. Arguments both for and against conventional superconductivity in this family, based on detailed experiments, continue to be presented[19,25-28]. Thus, future detailed characterization of the nickel-based oxypnictide superconductors BaNi$_2$P$_2$, La$_3$Ni$_4$P$_4$O$_2$ and LaNiPO will be of interest. The crystal structure of La$_3$Ni$_4$P$_4$O$_2$ is unique among the superconducting oxypnictides due to the asymmetric distribution of charge reservoir layers around the Ni$_2$P$_2$ plane. For the one copper oxide superconductor in which a similar structural asymmetry is known, Y$_2$Ba$_4$Cu$_7$O$_{15}$, unexpected behavior has been predicted[29]. Whether this asymmetry may be the reason why the $T_c$ in $La_3Ni_4P_4O_2$ is approximately half that of $BaNi_2P_2$ and LaNiPO is an issue that will require further study. The growth of single crystals will be required to provide more precise characterization of the normal state and superconducting properties.

**Acknowledgements**

This research at Los Alamos National Laboratory and Princeton University (grant DE-FG02-98ER45706) was supported by the US Department of Energy, Division of basic Energy Sciences. T.M.M. acknowledges support by the National Science Foundation.

**Table I**. Refined structural parameters for $La_3Ni_4P_4O_2$ at 298 K. Space group $I4/mmm$ (# 139), $a$ = 4.0107(1) Å, $c$ = 26.1811(9) Å. Calculated density 6.36 gm cm$^{-3}$. $U_{iso}$ = thermal vibration parameters in $10^{-2}$ Å$^2$. Figures of merit: goodness of fit ($\chi^2$) = 1.073, weighted profile residual ($R_{wp}$) = 4.87%, profile residual ($R_p$) = 3.98%, and residual on structure factors ($R(F^2)$) = 5.45%.

**$La_3Ni_4P_4O_2$ T = 298 K**

| Atom | $U_{iso}$ | Position | z |
|---|---|---|---|
| La1 | 0.88(5) | 2a (0,0,0) | |
| La2 | 0.88(5) | 4e (0,0,z) | 0.7013(1) |
| Ni | 1.15(4) | 8g (0,1/2,z) | 0.09278(7) |
| P1 | 0.84(7) | 4e (0,0,z) | 0.1335(2) |
| P2 | 0.84(7) | 4e (0,0,z) | 0.5493(2) |
| O | 1.2(1) | 4d (0,1/2,1/4) | |

**Figure Captions:**

**Figure 1.** (color on line) $La_3Ni_4P_4O_2$ is a member of a series of compounds built by inserting charged ions or layers between $M_2X_2$ sheets of the parent β-FeSe structure. All the oxypnictide superconductors are members of this family. The $La^{3+}$ and $O^{-2}$ ions are drawn with ionic radii and the M and X atoms are drawn with covalent radii.

**Figure 2.** (color on line) Superconducting transition measured in zero-field cooled (closed circles) and field cooled mode (open circles) under applied magnetic field 10 Oe. Inset of the figure represents magnetic susceptibility (ΔM/ΔH) in the wide temperature range. χ = ΔM/ΔH was calculated as the difference in magnetization ΔM between applied fields of 50 kOe and 40 kOe (see text).

**Figure 3.** (color on line) Magnetization (M) versus applied magnetic field (H) at constant temperature 1.9 K. The blue line corresponds to linear relation (~H) for H < 50 Oe. The inset shows deviation from a fitted linear dependence on H at a field H* (obtained from the fit between 5 and 50 Oe),

**Figure 4.** (color on line) Heat capacity data under 0 T and 9 T applied magnetic field. The green line represents a fit to the data (see text). The inset shows electronic specific heat ($C_{el.}/T$) in the low temperature range. The entropy conservation construction gives $\Delta C_{el.}/\gamma T_c$ = 1.25, underestimated due to the presence of the impurity phases.

**Figure 5.** (color on line) Temperature dependence of electrical resistance R(*T*)/R(300K). The upper inset shows the temperature dependence of the upper critical field ($H_{c2}$) from the midpoint of the superconducting transition (ρ(*T_c*)=1/2ρ_{res.}). Lower inset shows resistance data at low temperature, illustrating the changes in the superconducting transition with applied magnetic field. The presence of the LaNiPO minority phase results in a partial transition starting near 4.5 K that is suppressed quickly in the applied field.

10. $Ni_5P_4$ was synthesized from Ni powder (Alfa, 99.5%) and red P (Alfa 99.9%) by slow heating to 800 °C. La shavings (Johnson Matthey, 99.9%), $Ni_5P_4$, P, and dried $La_2O_3$ (Alfa 99.9%) were mixed, pelletized, placed in an alumina crucible in an evacuated quartz tube, and heated from 750 °C to 1050 °C over 1 hr. The sample was then reground, repressed, and reheated, with 2% excess P, at 1200 °C overnight. Finally, the sample was reground, repressed, and reheated at 1200 °C overnight (three times) in a tube backfilled with 1/3 atm. Ar.

11. Diffraction experiments were performed with neutrons of wavelength 1.5403 Å produced by a Cu(311) monochromator. Collimators with horizontal divergences of 15′, 20′ and 7' of arc were used before and after the monochromator, and after the sample, respectively. Data were collected in the $2\theta$ range of 3°–168° with a step size of 0.05°. The neutron scattering amplitudes used in the refinements were La = 0.824, Ni = 1.030, P = 0.513 and O = 0.581 x $10^{-12}$ cm.

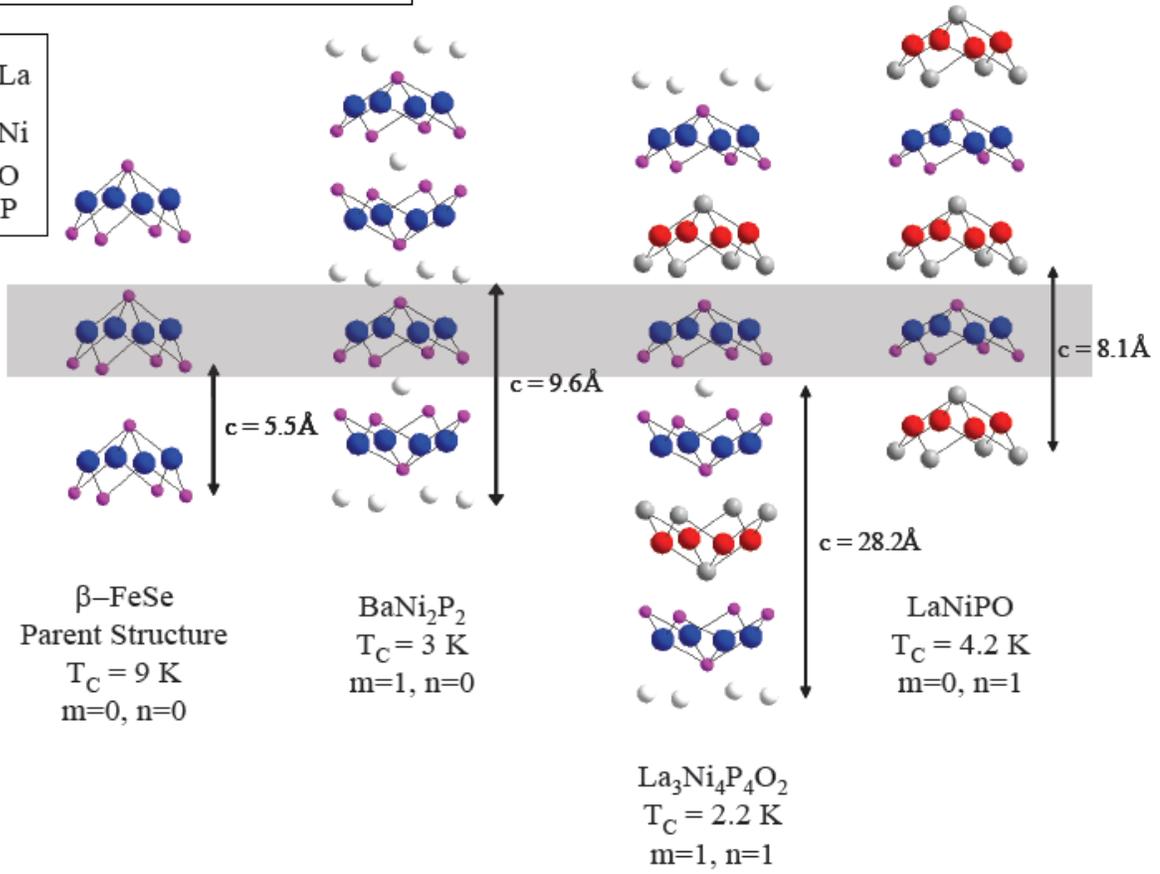

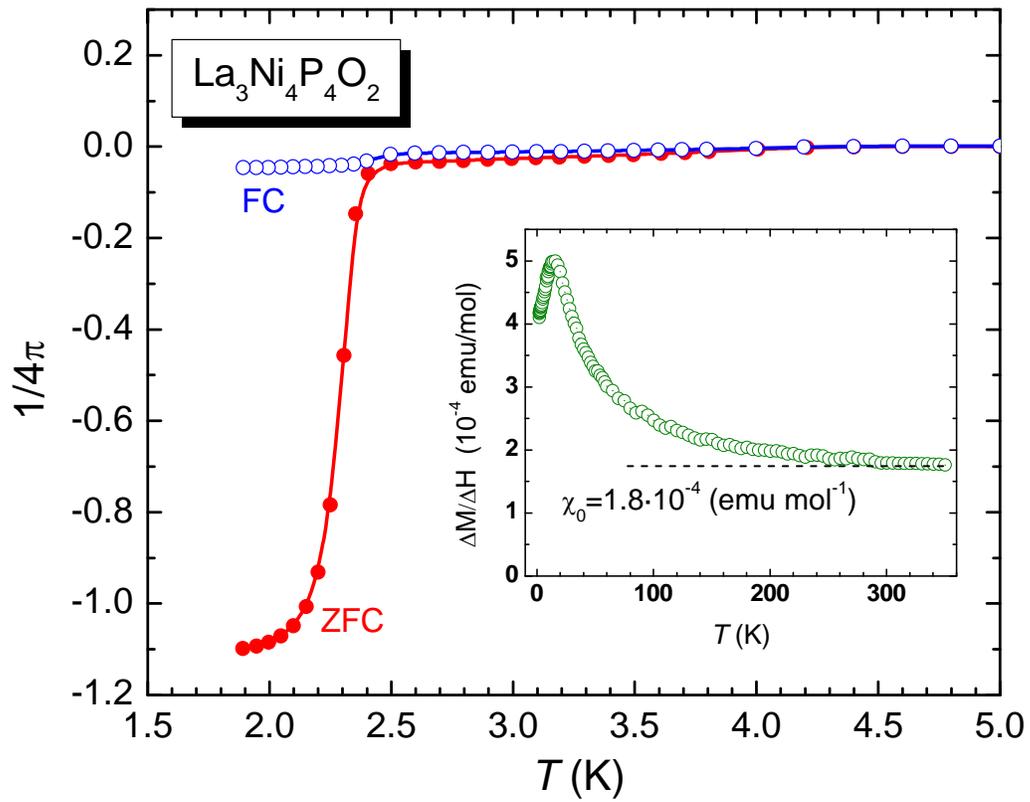

**Figure 2**

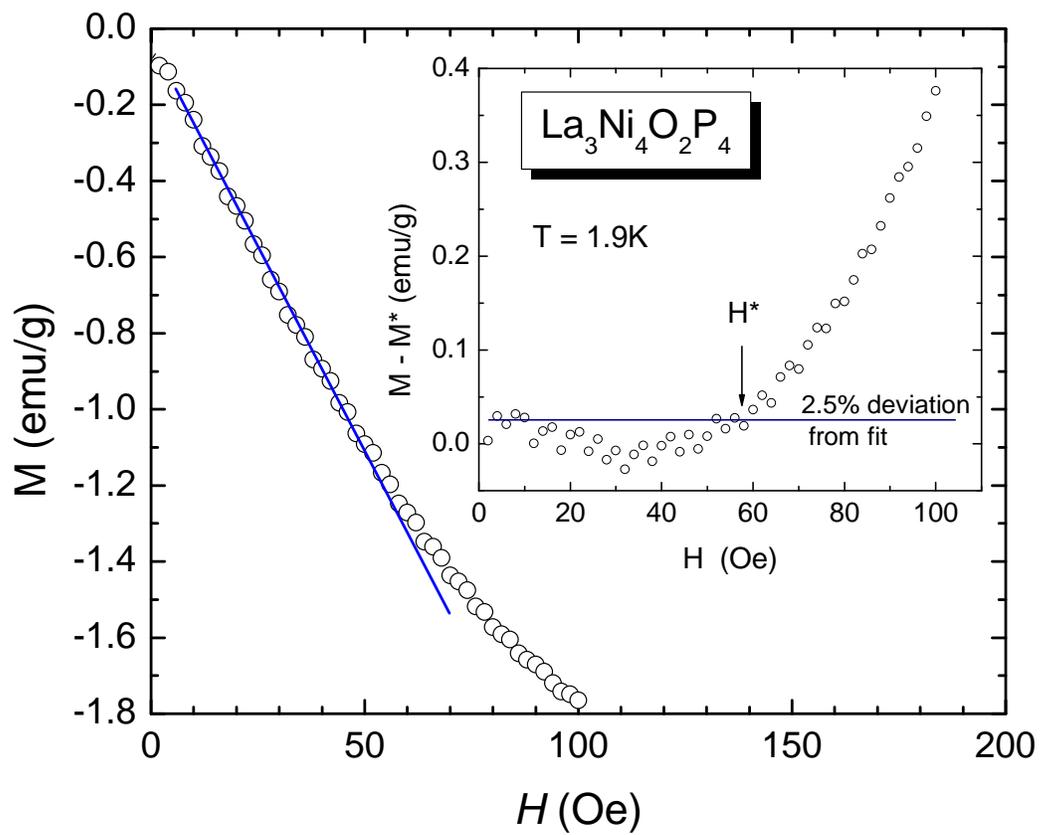

**Figure 3**

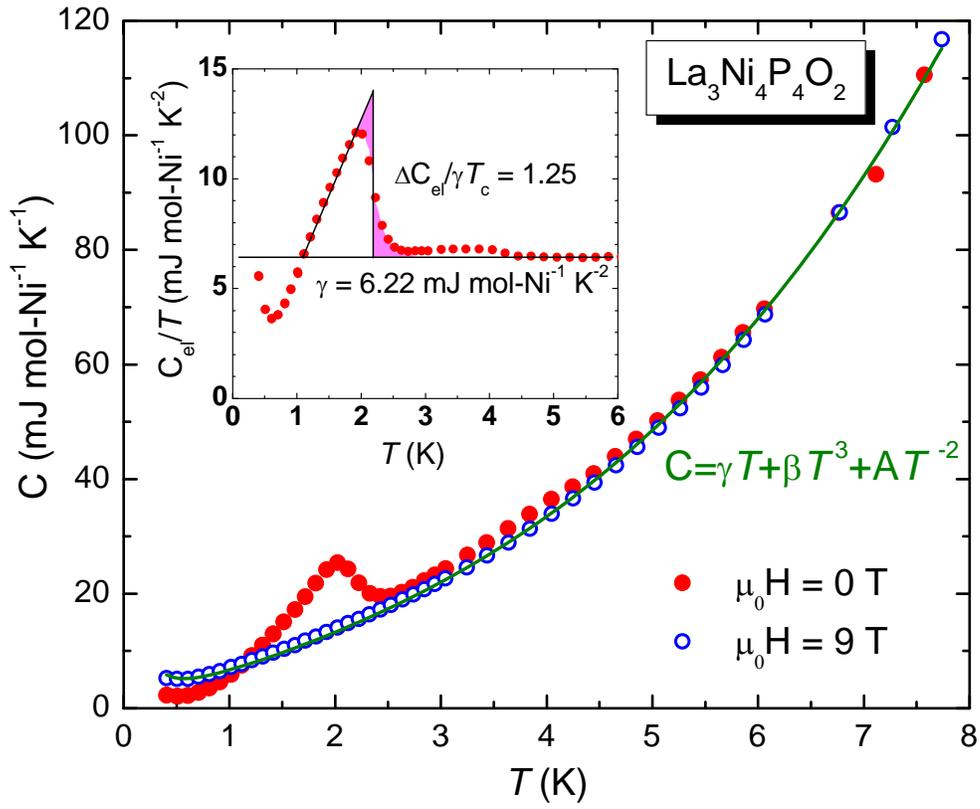

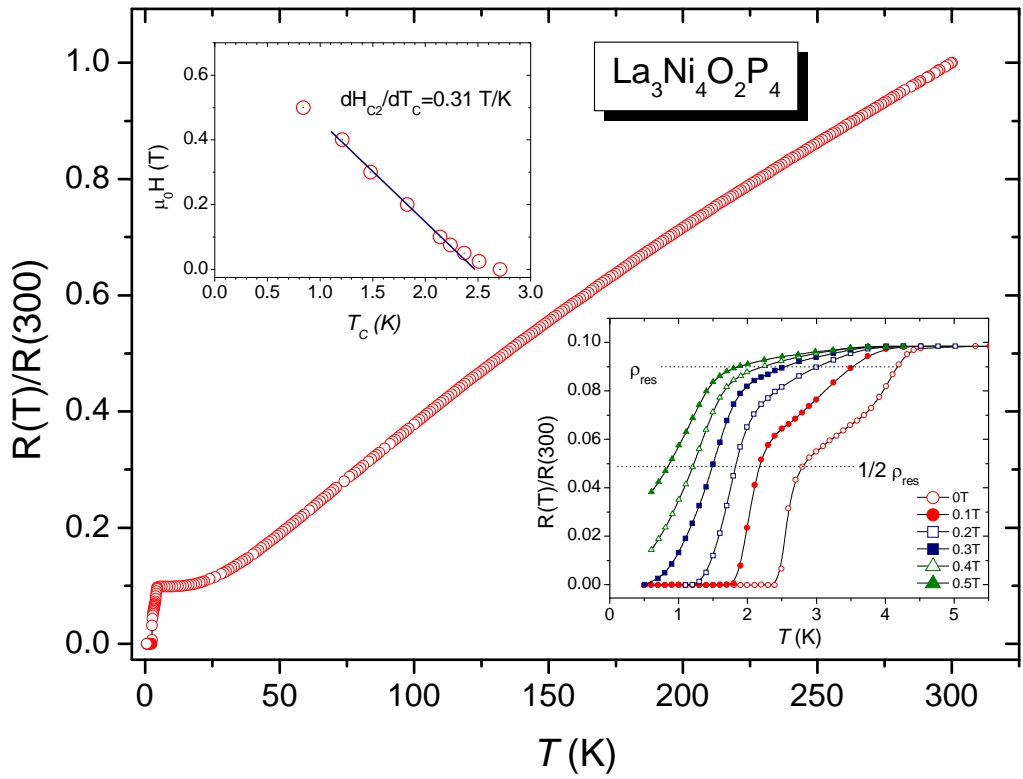

**Figure 5**

**Supplementary Figure.** Rietveld refinement of room temperature powder neutron diffraction data for $La_3Ni_4P_4O_2$. Upper part – crosses, observed data, solid lines calculated intensities. The lower part shows on the same scale the differences between the observed and calculated pattern The blue tick marks correspond to $La_3Ni_4P_4O_2$, with the black and green sets referring to the impurities LaNiPO (10 %) and $LaNi_5P_3$ (6 %) respectively. The quality of the fit is excellent and confirms the stoichiometry and crystal structure of the title compound.

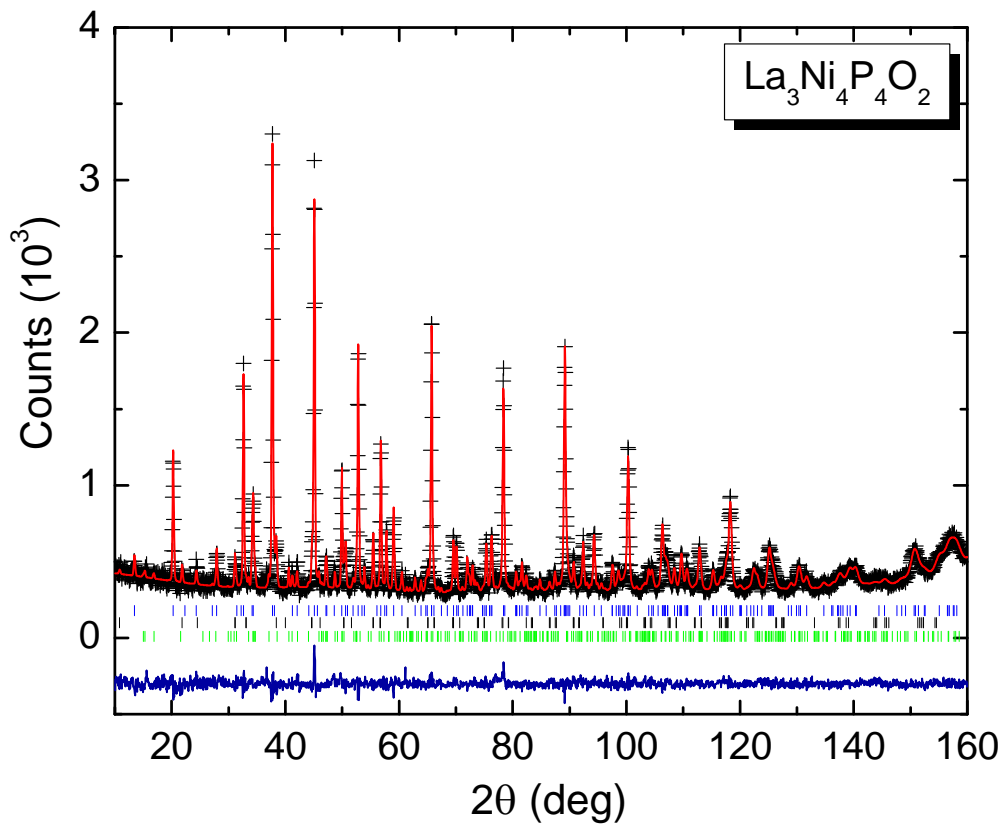